\documentclass[a4paper]{article}

\usepackage{INTERSPEECH2021}
\pdfoutput=1

\title{Basis-MelGAN: Efficient Neural Vocoder Based on Audio Decomposition}
\name{Zhengxi Liu$^1$, Yanmin Qian$^{2,\dagger}$\thanks{$^{\dagger}$Yanmin Qian is the corresponding author}}
\address{
  $^1$Sun Yat-Sen University\\
  $^2$MoE Key Lab of Artificial Intelligence, AI Institute\\
  X-LANCE Lab, Department of Computer Science and Engineering\\
  Shanghai Jiao Tong University, Shanghai, China}
\email{liuzhx33@mail2.sysu.edu.cn, yanminqian@sjtu.edu.cn}

\begin{document}

\maketitle
\begin{abstract}
Recent studies have shown that neural vocoders based on generative adversarial network (GAN) can generate audios with high quality. While GAN based neural vocoders have shown to be computationally much more efficient than those based on autoregressive predictions, the real-time generation of the highest quality audio on CPU is still a very challenging task. One major computation of all GAN-based neural vocoders comes from the stacked upsampling layers, which were designed to match the length of the waveform's length of output and temporal resolution. Meanwhile, the computational complexity of upsampling networks is closely correlated with the numbers of samples generated for each window. To reduce the computation of upsampling layers, we propose a new GAN based neural vocoder called Basis-MelGAN where the raw audio samples are decomposed with a learned basis and their associated weights. As the prediction targets of Basis-MelGAN are the weight values associated with each learned basis instead of the raw audio samples, the upsampling layers in Basis-MelGAN can be designed with much simpler networks. 
Compared with other GAN based neural vocoders, the proposed Basis-MelGAN could produce comparable high-quality audio but significantly reduced computational complexity from HiFi-GAN V1's 17.74 GFLOPs to 7.95 GFLOPs.


\end{abstract}

\section{Introduction}

Neural vocoders have made extraordinary success in recent studies. To date, the neural vocoders that generate the highest quality audios are based on autoregressive prediction, such as WaveNet \cite{DBLP:journals/corr/OordDZSVGKSK16} and WaveRNN \cite{pmlr-v80-kalchbrenner18a}. However, these autoregressive-based models have been suffered from low inference speed due to their high computational complexity and the difficulties of parallelization.
To achieve the parallelization on GPU, several non-autoregressive based neural vocoders, such as Parallel WaveNet \cite{pmlr-v80-oord18a}, WaveGlow \cite{8683143} and ClariNet \cite{DBLP:journals/corr/abs-1807-07281}, have been proposed.
While these neural vocoders based on parallelization significantly improve the inference speed, these improvements are only applicable when the model inferences on GPU. More importantly, the total computational complexity does not reduce with parallelization. 
To reduce the total computational complexity of neural vocoders, several approaches such as LPCNet \cite{8682804}, Multi-Band WaveRNN \cite{DBLP:journals/corr/abs-1909-01700} and FeatherWave \cite{tian2020featherwave} are proposed to utilize existing signal processing techniques to simplify the model complexity. However, even with these improvements, the computational complexity of neural vocoders are remained very high.

Recently, several GAN-based non-autoregressive models are proposed which produce high-quality audio with significantly less computation than their alternatives. For example, MelGAN \cite{NEURIPS2019_6804c9bc} and HiFi-GAN \cite{NEURIPS2020_c5d73680} can even produce audio in real-time on CPU with relative high quality. GAN-based neural vocoders' major strength is that it can generate a window of audio samples (e.g., 256) at every inference step, which is a significant improvement in computation compared to previous neural vocoders where only a single audio sample can be generated at every inference step.

While GAN-based neural vocoders have shown to be computationally much more efficient than these based on autoregressive predictions, the real-time generation of the highest quality audio on the CPU is still a very challenging task. Some of the neural vocoders based on GAN can produce speech in real-time on CPU, but sample quality is relatively lower. For example, HiFi-GAN V1 can produce high-quality speech, but it cannot infer in real-time on some middle or low-end devices. Therefore, the computation of GAN-based neural vocoders needs to be further reduced.


One major computation of all GAN-based neural vocoders comes from the stacked upsampling layers, which were designed to match the waveform's length of output and temporal resolution. As the complexity of upsampling layers is closely correlated with the number of samples in each window to be predicted, a more effective representation of the signal in each window can reduce the number target dimension, reducing the complexity of upsampling layers. Motivated by this, we propose to represent audio signals more compactly and efficiently to mitigate upsampling networks' complexity. Specifically, we decompose audio signals with a learned basis and their associated weights. With this decomposition, audio signals can be efficiently represented with a nonnegative weighted sum of the N basis matrix. Since the basis is fixed, only the weights associated with each basis need to be predicted. As the number of bases is much smaller than the raw audio waveform, the target output has much less dimension, which means a much simple upsampling network is required for matching the output dimension.

There are several existing methods in terms of audio decomposition, especially in the related field of blind audio separation. Traditionally, the audio signal decomposition can be achieved with independent component analysis (ICA) \cite{4815260} and time-domain nonnegative matrix factorization (NMF) \cite{4685898}. Recent studies have explored the use of deep learning for audio decomposition and made great success in the field of audio separation \cite{7310882, 7952123}. The deep learning based audio decomposition is a more data-driven representation that can be learned to minimize the audio reconstruction loss. This deep learning based audio decomposition has another advantage in that it can be implicitly incorporated into other network architectures and be updated jointly with other objectives. In this study, we choose the TasNet network \cite{8462116}, which was initially proposed for the task of audio separation, to learn the basis of audio signals for decomposition. One major advantage of using TasNet for learning audio decomposition is that the learned basis has better generality than using a simple audio reconstruction as an objective.

Finally, to enhance Basis-MelGAN's modelling capabilities of time-frequency characteristics, we use multi-resolution STFT discriminator. The multi-resolution STFT discriminator show better performance than the multi-period discriminator used in HiFi-GAN and more efficient since the input to the discriminator is spectrogram instead of the raw waveform during adversarial training. \footnote{Audio samples can be found in https://blog.xcmyz.xyz/demo/ and code can be found in https://github.com/xcmyz/FastVocoder.}

\section{Proposed Method}

The proposed model consists of two parts, TasNet and Basis-MelGAN. We first train a TasNet model to get basis matrix that will be used for audio decomposition in Basis-MelGAN model. 
The basis matrix learned from TasNet will be used as frozen parameters in Basis-MelGAN generator to train the model. We will first show the design and training of TasNet. The details of the Basis-MelGAN will be introduced in the following sections.

\subsection{TasNet}

\begin{figure}[t]
  \centering
  \includegraphics[width=\linewidth]{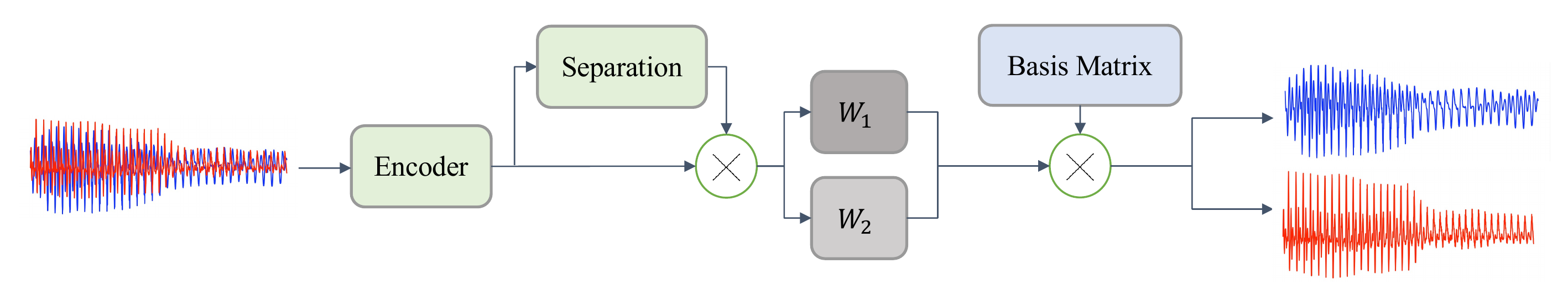}
  \caption{Model structure of TasNet.}
  \label{fig:tasnet}
\end{figure}

TasNet is a single-channel speech separation model. The input of TasNet is mixture of speech from different sources $ x \in \mathbb{R} ^ {1 \times T} $ of $ C $ sources $ s_1,\cdots,s_c \in \mathbb{R} ^ {1 \times T} $, where the purpose of TasNet model is to estimate $ s_1,\cdots,s_c $, from mixture speech of $ x $. 
The model composes of three parts as in Figure~\ref{fig:tasnet}. These are encoder network, separation network, and a basis matrix that can be learned jointly with other networks. The Encoder network consists of a 1-D convolution layer and a ReLU activation function. 
The separation module is a fully convolutional network that consists of stacked 1-D dilated convolutional blocks. The output of encoder $ \hat{\mathbf{W}} $ is used as the input of the separation module. Both the encoder and separation network is the same with Conv-TasNet \cite{8707065}. The main difference in this work is that we uses two masks $ \mathbf{M}_1, \mathbf{M}_2 \in \mathbb{R} ^ {N \times \hat{T}} $ for two mixed audio shown as figure~\ref{fig:tasnet}. Since the objective of training TasNet is to obtain the basis of audio for decomposition, we do not design the network's output to separate audios from different sources as in the original TasNet framework. Instead, we add random noises to each audio input, and the objective is to reconstruct original input audio. Therefore, the input to TasNet is a mixture of clean speech and noise, and the model is trained to separate them with two masks of $ \mathbf{M}_1, \mathbf{M}_2$. Specifically, the weight matrix associated with clean speech and noise are obtained by 

\begin{equation}
\mathbf{W}_i = \hat{\mathbf{W}} \cdot \mathbf{M}_i
\end{equation}

where $ i \in \{1,2\}$, represents the clean speech and noise.
Then, the separated $ \mathbf{W}_i $ are multiplied with the basis matrix to obtain the separated speech and noise signals $y_i$ via

\begin{equation}
y_i = \mathbf{B} \cdot \mathbf{W}_i
\end{equation}



$\mathbf{B}$ represents the basis matrix of size $[m, n]$ with $m$ the length of each audio window to be decomposed, and $n$ represents the number of basis for decomposition. $m$ is normally much smaller than $n$ to be able to represent raw audio more effectively. In this work, we use the number of basis used as 256, and the window length of audio is 32. Therefore, the basis matrix is [32, 256], and each window of audio with a length of [1, 32] can be decomposed and represented with associated weight values of [1, 256].
The training objective uses SI-SNR as in the original TasNet paper.

\subsection{Basis-MelGAN}

\begin{figure}[t]
  \centering
  \includegraphics[width=0.8\linewidth]{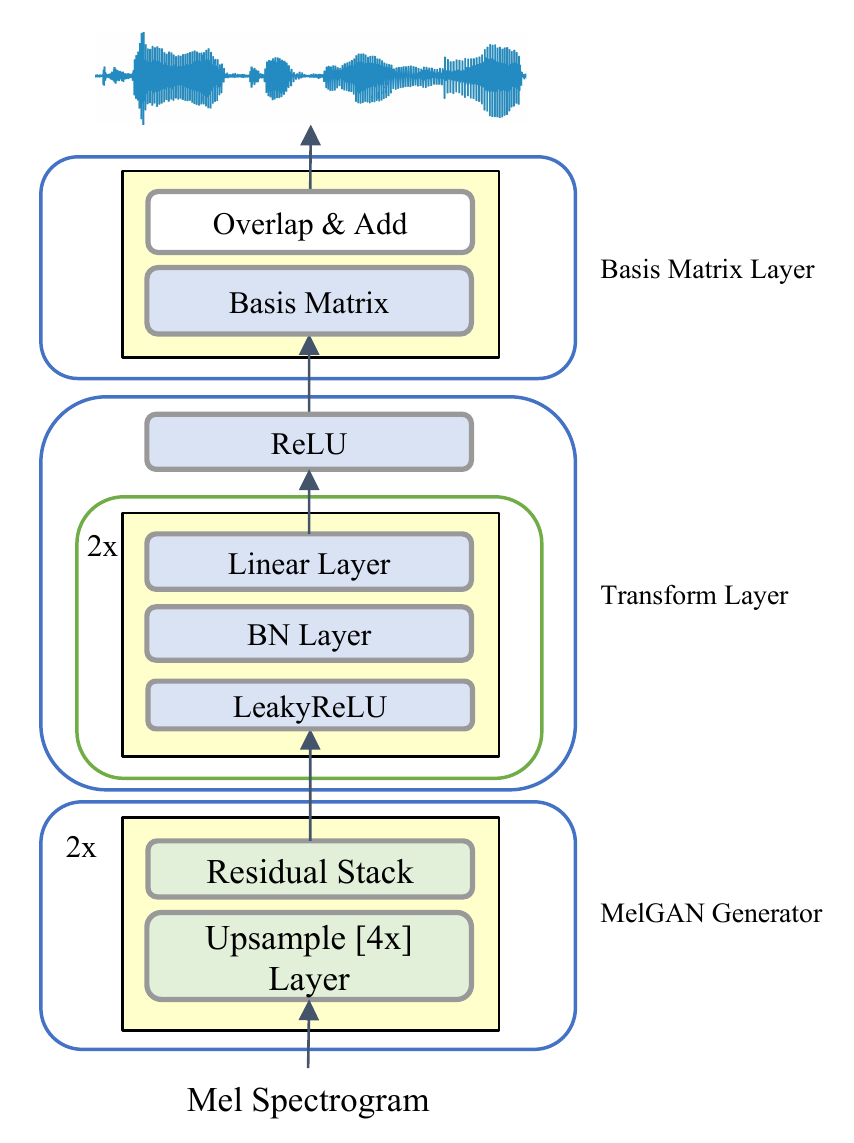}
  \caption{Model structure of Basis-MelGAN.}
  \label{fig:model}
\end{figure}

The proposed Basis-MelGAN consists of three parts: MelGAN Generator, Transform layer, and Basis Matrix learned from TasNet. It takes mel-spectrogram as input and output audio waveform as shown in Figure~\ref{fig:model}.

The generator of Basis-MelGAN shares the same structure as MelGAN. It is a fully convolutional network consists of a stack of transposed convolutional layers to upsample the input mel spectrogram to have the same resolution of time-domain audio. Each transposed convolutional layer is followed by a stack of residual blocks with dilated convolutions. 

The transform layer is a linear feed-forward network, which consists of two stacks of linear layer with a leaky ReLU \cite{10.5555/3104322.3104425}, a batch normalization layer \cite{pmlr-v37-ioffe15} and a linear network. A ReLU activation function is added to produce nonnegative weight.

Finally, the basis matrix layer is the same as TasNet mentioned in section 2.1. It shares the same parameter of the basis matrix with the TasNet basis matrix. Primarily, we train Basis-MelGAN by frozen the parameter of the basis matrix because it shows the best performance in this way. The model can be in convergence with unfrozen basis matrix in random initialization, but which performance is not good as freezing basis matrix with TasNet parameter.

\subsection{STFT Discriminator}

\begin{figure}[t]
  \centering
  \includegraphics[width=\linewidth]{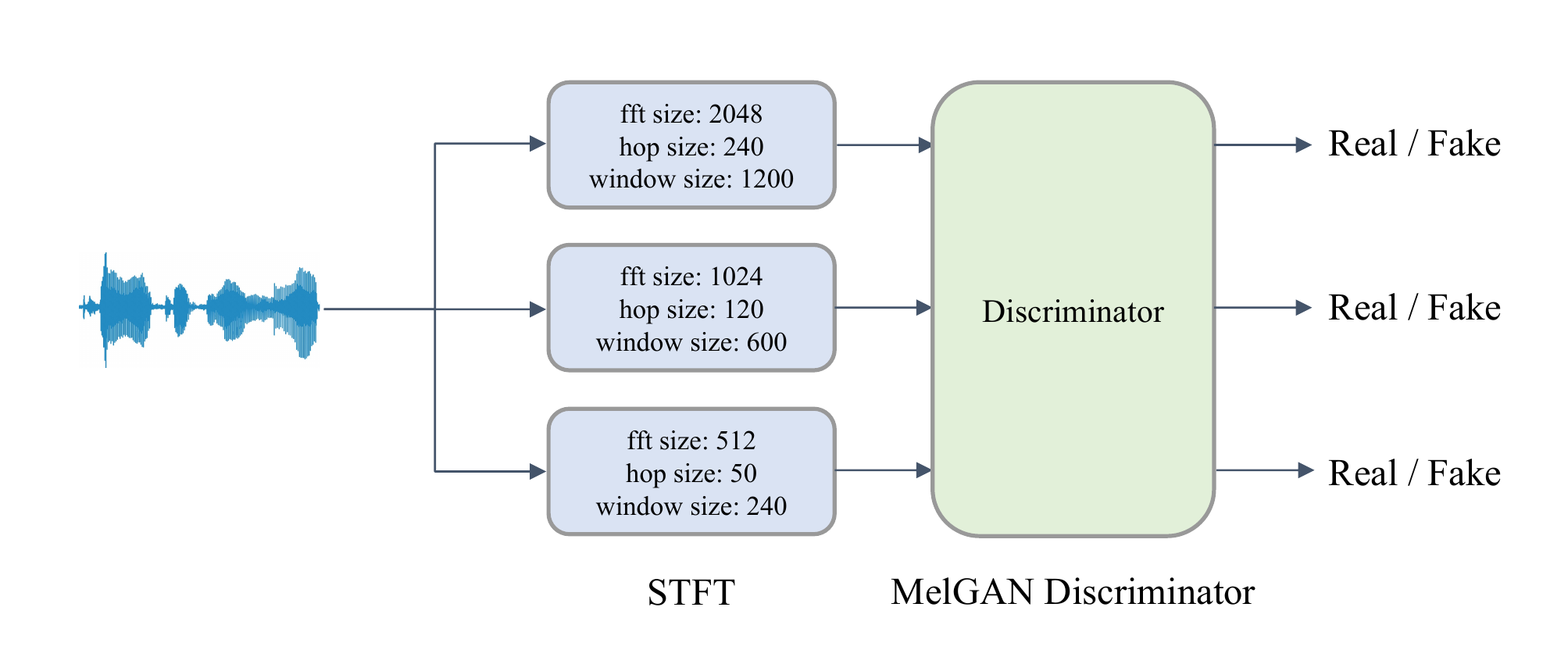}
  \caption{Model structure of multi-resolution STFT discriminator.}
  \label{fig:discriminator}
\end{figure}

We use a new discriminator called multi-resolution STFT discriminator (MFD), which is a discriminator with input of spectrogram. It has the same architecture as a multi-scale discriminator (MSD) in MelGAN and the same multi-resolution settings as multi-resolution STFT loss in Parallel WaveGAN \cite{9053795} shown as figure~\ref{fig:discriminator}. As discriminator is a more powerful criterion than L1 loss function \cite{NIPS2014_5ca3e9b1}, it helps generator learn time-frequency characteristics more efficiently so that generator can generate high quality audio with more details in frequency domain. We combine MFD with MSD in adversarial training. This combination makes the generator learn both time-frequency characteristics and the distribution of the speech waveform. Besides, for ablation study, we also train generators with the same multi-discriminator architecture as HiFI-GAN, i.e., multi-period discriminator (MPD) and MSD. We find the combination of MFD and MSD shows the best performance, which achieves a higher MOS score of 4.21 than 4.13 of the combination of MPD and MSD. In addition, MFD is much faster than MPD during training. The combination of MFD and MSD only spends $ 1/4 $ time to reach the same step as the combination of MPD and MSD since the input to the discriminator is spectrogram instead of the raw waveform during adversarial training. \footnote{At the time of preparing this paper, we became aware that a preprint paper Universal MelGAN \cite{jang2021universal} has the similar idea to improve the efficiency of adversarial training for GAN based vocoder, which is using spectrogram as the input of discriminator instead of the raw waveform. Our work is independently developed and the fact that many design choices are completely different.}

\subsection{Loss Function}

We use four different loss functions to train the Basis-MelGAN generator, which are weight loss $ L_{weight} $, multi-resolution STFT loss $ L_{stft} $, multi-scale adversarial loss $ L_{adv\_s} $ and multi-resolution STFT adversarial loss $ L_{adv\_f} $. We do not use feature matching loss in MelGAN, since the model can not converge when add this loss. For weight loss $ L_{weight} $, we minimize the $ \ell1 $ norm between the target weight $ \mathbf{W} $ from TasNet and the predicted weight $ \mathbf{\overline{W}} $ from the Basis-MelGAN generator, where:

\begin{equation}
L_{weight} = \left\| \mathbf{W} - \mathbf{\overline{W}} \right\|_1
\end{equation}

For single STFT loss $ L_{single\_stft} $, we minimize the spectral convergence $ L_{sc} $ and log STFT magnitude $ L_{mg} $ between the target waveform $ y $ from TasNet and the predicted waveform $ \overline{y} $ from the Basis-MelGAN generator. Hence the objective of $ L_{sc} $ and $ L_{mg} $ becomes ($ \lvert stft(\cdot) \rvert $ indicates the STFT function to compute magnitudes and $ N $ is the number of elements in the magnitude):

\begin{gather}
L_{sc} = \frac{\left\| \lvert stft(y)\rvert - \lvert stft(\overline{y}) \rvert \right\|_F}{\left\| \lvert stft(y) \rvert \right\|_F}  \\
L_{mg} = \frac{1}{N}\left\| log\lvert stft(y) \rvert - \log\lvert stft(\overline{y}) \rvert \right\|_1 \\
L_{single\_stft} = L_{sc} + L_{mg}
\end{gather}

For the multi-resolution STFT loss function, we use M single STFT loss functions with different STFT parameters (i.e., FFT size, window size and hop size). Therefore, the multi-resolution STFT loss function is shown as follow:

\begin{equation}
L_{mr\_stft} = \frac{1}{M} \sum_{m=1}^{M} L_{single\_stft}^{m}
\end{equation}

For multi-scale adversarial loss $ L_{adv\_s} $ and multi-resolution STFT adversarial loss $ L_{adv\_f} $, we minimize the binary cross-entropy between the output of discriminator passed by target waveform $ y $ and the output of discriminator passed by Basis-MelGAN output $ \overline{y} $, where:

\begin{gather}
L_{adv\_s} = \frac{1}{N_s}\operatorname{BCELoss}(\operatorname{MSD}(y), \operatorname{MSD}(\overline{y})) \\
L_{adv\_f} = \frac{1}{N_f}\operatorname{BCELoss}(\operatorname{MFD}(y), \operatorname{MFD}(\overline{y}))
\end{gather}

$ N_s $, $ N_f $ indicate the number of single discriminator in multi-scale discriminator and multi-resolution STFT discriminator. Thus, the total loss $ L_G $ for Basis-MelGAN generator is shown as following:

\begin{equation}
L_G = L_{sc} + L_{mg} + L_{adv\_s} + L_{adv\_f}
\end{equation}

For training multi-scale discriminator and multi-resolution STFT discriminator, we minimize the binary cross-entropy $ L_{dis\_real} $ between the output of discriminator passed by target waveform $ y $ and real label, and the binary cross-entropy $ L_{dis\_fake} $ between the output of discriminator passed by Basis-MelGAN output $ \overline{y} $ and fake label, where:

\begin{gather}
L_{dis\_real} = \frac{1}{N}\operatorname{BCELoss}(\operatorname{D}(y), 1) \\
L_{dis\_fake} = \frac{1}{N}\operatorname{BCELoss}(\operatorname{D}(\overline{y}), 0)
\end{gather}

$ N $ indicates the number of single discriminators in multi-scale discriminator or multi-resolution STFT discriminator. D indicates MSD or MFD. $ 1 $ indicates real label and $ 0 $ indicates fake label.

\section{Experiments}

We compare our model with the state-of-the-art GAN based neural vocoder model HiFi-GAN and a very fast GAN based neural vocoder model Multi-Band MelGAN \cite{yang2020multiband}.

\subsection{Dataset}

For experiments, we use an open-source single-speaker dataset LJSpeech, which contains 13,100 short audio clips of a single speaker reading passages from 7 non-fiction books. We leave out 100 sentences from the corpus for testing. We use the STFT same settings as \cite{NEURIPS2020_c5d73680} for fair comparison.

\subsection{Model details}

\subsubsection{TasNet}
We use the same structure as Conv-TasNet and change the hyper-parameter of Conv-TasNet to adapt our model.
We train the model to separate mixture which are mixed by waveform from LJSpeech and random noise sampled from normal distribution with 0 mean and 0.03125 std. After training the TasNet for 300k steps, we get a basis matrix, and it will be used as frozen parameters of Basis-MelGAN.

\subsubsection{Basis-MelGAN}
The network structure of Basis-MelGAN is similar to MelGAN and we only use two 4x upsampling layers instead of four upsampling layers with [8, 8, 2, 2] upsampling scales in MelGAN. Besides, we experiment with a light footprint version model for better inference speed, which reaches the fastest inference speed among our testing models.

\subsection{Training details}

We first train the Basis-MelGAN generator with weight loss and multi-resolution STFT loss for 300k steps. We start to use adversarial training from 300k to 1M steps, and during adversarial training, we do not use weight loss since we find it achieves better performance. We use Adam optimizer \cite{DBLP:journals/corr/KingmaB14} with initial learning rate as 0.001 for generator and 0.0005 for discriminator. We train our model on Nvidia V100 (16G). 

\subsection{Evaluation}


%

\begin{table}[t]
  \caption{MOS scores of each model.}
  \label{tab:mos}
  \centering
  \begin{tabular}{lcc}
    \toprule
    \textbf{Model Name} & \textbf{MOS} & \textbf{95\% CI} \\
    \midrule
    Ground Truth & 4.56 & $ \pm $0.09 \\
    \midrule
    Basis-MelGAN (Large) & 4.21 & $ \pm $0.10 \\
    Basis-MelGAN (Light) & 4.15 & $ \pm $0.11 \\
    \midrule
    HiFi-GAN V1 & 4.25 & $ \pm $0.11 \\
    HiFi-GAN V2 & 4.18 & $ \pm $0.10 \\
    HiFi-GAN V3 & 4.10 & $ \pm $0.11 \\
    Multi-Band MelGAN & 3.97 & $ \pm $0.10 \\
    \bottomrule
  \end{tabular}
\end{table}

\subsubsection{Quality}

Mean Opinion Score (MOS) of the naturalness of generated speech utterances are rated by human subjects who participated in the listening tests. We use ground-truth mel spectrogram as input and evaluate the quality of audio generated by testing models. The results are shown in Table~\ref{tab:mos}. ``Basis-MelGAN (Large)'' means the original Basis-MelGAN and ``Basis-MelGAN (Light)'' means the light footprint version Basis-MelGAN. Remarkably, Basis-MelGAN (Large) achieves the MOS score of 4.21 with a tiny gap of 0.04 compared to HiFi-GAN V1, but Basis-MelGAN (Large) is 2.6 times faster than HiFi-GAN V1. Besides, for the ablation study, we train Basis-MelGAN without frozen basis matrix, which reaches the MOS score of 3.93, which is much lower than Basis-MelGAN with frozen basis matrix. This ablation study shows the necessity of use a frozen basis matrix from TasNet to improve the audio quality.

\begin{table}[t]
  \caption{MOS scores of end-to-end speech synthesis.}
  \label{tab:mos_e2e}
  \centering
  \begin{tabular}{lcc}
    \toprule
    \textbf{Model Name} & \textbf{MOS} & \textbf{95\% CI} \\
    \midrule
    Ground Truth & 4.51 & $ \pm $0.07 \\
    \midrule
    Griffin Lim & 3.12 & $ \pm $0.06 \\
    \midrule
    Basis-MelGAN (Large) & 4.10 & $ \pm $0.08 \\
    Basis-MelGAN (Light) & 4.02 & $ \pm $0.07 \\
    \midrule
    HiFi-GAN V1 & 4.12 & $ \pm $0.07 \\
    HiFi-GAN V2 & 4.03 & $ \pm $0.07 \\
    HiFi-GAN V3 & 3.99 & $ \pm $0.08 \\
    Multi-Band MelGAN & 3.82 & $ \pm $0.07 \\
    \bottomrule
  \end{tabular}
\end{table}

We also examine the proposed models' effectiveness when applied to an end-to-end speech synthesis pipeline, which is a acoustic model for text to mel spectrogram and a neural vocoder for mel spectrogram to waveform. We use Tacotron2 \cite{8461368} as an acoustic model and make fine-tuning training with predicted mel spectrogram of Tacotron2 on all testing models. The MOS scores are listed in Table~\ref{tab:mos_e2e}. Our models show a robust ability to adapt to end-to-end speech synthesis.

\subsubsection{Inference speed}

\begin{table}[t]
  \caption{RTF of models}
  \label{tab:rtf}
  \centering
  \begin{tabular}{lccc}
    \toprule
    \textbf{Model Name} & \textbf{Low} & \textbf{High} & \textbf{Para (M)} \\
    \midrule
    Basis-MelGAN (Light) & 0.1460 & 0.0100 & 3.30 \\
    Basis-MelGAN (Large) & 0.6668 & 0.0395 & 15.90 \\
    HiFi-GAN V1 & 1.8786 & 0.1033 & 13.92 \\
    HiFi-GAN V2 & 0.1960 & 0.0303 & 0.92 \\
    HiFi-GAN V3 & 0.1977 & 0.0213 & 1.46 \\
    MB MelGAN & 0.1351 & 0.0175 & 2.53 \\
    \bottomrule
  \end{tabular}
\end{table}

We test the real-time factor (RTF) of neural vocoder models on a low-end platform and high-end platform, shown as Table~\ref{tab:rtf}. Low end platform is single-core AMD EPYC 7551 (2.0 GHz, 2GB RAM) and high end platform is 8 core Intel(R) Xeon(R) Gold 6146 (16GB RAM). Basis-MelGAN (Light) reaches the highest speed, which is even faster than Multi-Band MelGAN. Basis-MelGAN (Large) reaches 0.67 RTF on low-end platform. Meanwhile, HiFi-GAN V1 can not be in real-time.

\section{Conclusions}

We have introduced a GAN-based neural vocoder model, which has a novel architecture, using TasNet basis matrix as a part of the model, and have shown this design makes the improvement of inference speed and audio quality. Our work demonstrates the feasibility of using audio decomposition in neural vocoder. We hope there will be more deep learning-based audio decomposition techniques used in speech synthesis to accelerate inference speed and improve audio quality.

\bibliographystyle{IEEEtran}

\bibliography{template}

\begin{thebibliography}{10}
\providecommand{\url}[1]{#1}
\csname url@samestyle\endcsname
\providecommand{\newblock}{\relax}
\providecommand{\bibinfo}[2]{#2}
\providecommand{\BIBentrySTDinterwordspacing}{\spaceskip=0pt\relax}
\providecommand{\BIBentryALTinterwordstretchfactor}{4}
\providecommand{\BIBentryALTinterwordspacing}{\spaceskip=\fontdimen2\font plus
\BIBentryALTinterwordstretchfactor\fontdimen3\font minus
  \fontdimen4\font\relax}
\providecommand{\BIBforeignlanguage}[2]{{%
\expandafter\ifx\csname l@#1\endcsname\relax
\typeout{** WARNING: IEEEtran.bst: No hyphenation pattern has been}%
\typeout{** loaded for the language `#1'. Using the pattern for}%
\typeout{** the default language instead.}%
\else
\language=\csname l@#1\endcsname
\fi
#2}}
\providecommand{\BIBdecl}{\relax}
\BIBdecl

\bibitem{DBLP:journals/corr/OordDZSVGKSK16}
A.~van~den Oord, S.~Dieleman, H.~Zen, K.~Simonyan, O.~Vinyals, A.~Graves,
  N.~Kalchbrenner, A.~W. Senior, and K.~Kavukcuoglu, ``Wavenet: {A} generative
  model for raw audio,'' \emph{CoRR}, vol. abs/1609.03499, 2016.

\bibitem{pmlr-v80-kalchbrenner18a}
N.~Kalchbrenner, E.~Elsen, K.~Simonyan, S.~Noury, N.~Casagrande, E.~Lockhart,
  F.~Stimberg, A.~van~den Oord, S.~Dieleman, and K.~Kavukcuoglu, ``Efficient
  neural audio synthesis,'' in \emph{Proceedings of the 35th International
  Conference on Machine Learning}, ser. Proceedings of Machine Learning
  Research, J.~Dy and A.~Krause, Eds., vol.~80.\hskip 1em plus 0.5em minus
  0.4em\relax PMLR, 10--15 Jul 2018, pp. 2410--2419.

\bibitem{pmlr-v80-oord18a}
A.~van~den Oord, Y.~Li, I.~Babuschkin, K.~Simonyan, O.~Vinyals, K.~Kavukcuoglu,
  G.~van~den Driessche, E.~Lockhart, L.~Cobo, F.~Stimberg, N.~Casagrande,
  D.~Grewe, S.~Noury, S.~Dieleman, E.~Elsen, N.~Kalchbrenner, H.~Zen,
  A.~Graves, H.~King, T.~Walters, D.~Belov, and D.~Hassabis, ``Parallel
  {W}ave{N}et: Fast high-fidelity speech synthesis,'' in \emph{Proceedings of
  the 35th International Conference on Machine Learning}, ser. Proceedings of
  Machine Learning Research, J.~Dy and A.~Krause, Eds., vol.~80.\hskip 1em plus
  0.5em minus 0.4em\relax PMLR, 10--15 Jul 2018, pp. 3918--3926.

\bibitem{8683143}
R.~{Prenger}, R.~{Valle}, and B.~{Catanzaro}, ``Waveglow: A flow-based
  generative network for speech synthesis,'' in \emph{ICASSP 2019 - 2019 IEEE
  International Conference on Acoustics, Speech and Signal Processing
  (ICASSP)}, 2019, pp. 3617--3621.

\bibitem{DBLP:journals/corr/abs-1807-07281}
W.~Ping, K.~Peng, and J.~Chen, ``Clarinet: Parallel wave generation in
  end-to-end text-to-speech,'' \emph{CoRR}, vol. abs/1807.07281, 2018.

\bibitem{8682804}
J.~{Valin} and J.~{Skoglund}, ``Lpcnet: Improving neural speech synthesis
  through linear prediction,'' in \emph{ICASSP 2019 - 2019 IEEE International
  Conference on Acoustics, Speech and Signal Processing (ICASSP)}, 2019, pp.
  5891--5895.

\bibitem{DBLP:journals/corr/abs-1909-01700}
C.~Yu, H.~Lu, N.~Hu, M.~Yu, C.~Weng, K.~Xu, P.~Liu, D.~Tuo, S.~Kang, G.~Lei,
  D.~Su, and D.~Yu, ``Durian: Duration informed attention network for
  multimodal synthesis,'' \emph{CoRR}, vol. abs/1909.01700, 2019.

\bibitem{tian2020featherwave}
Q.~Tian, Z.~Zhang, H.~Lu, L.-H. Chen, and S.~Liu, ``Featherwave: An efficient
  high-fidelity neural vocoder with multi-band linear prediction,'' 2020.

\bibitem{NEURIPS2019_6804c9bc}
K.~Kumar, R.~Kumar, T.~de~Boissiere, L.~Gestin, W.~Z. Teoh, J.~Sotelo,
  A.~de~Br\'{e}bisson, Y.~Bengio, and A.~C. Courville, ``Melgan: Generative
  adversarial networks for conditional waveform synthesis,'' in \emph{Advances
  in Neural Information Processing Systems}, H.~Wallach, H.~Larochelle,
  A.~Beygelzimer, F.~d\textquotesingle Alch\'{e}-Buc, E.~Fox, and R.~Garnett,
  Eds., vol.~32.\hskip 1em plus 0.5em minus 0.4em\relax Curran Associates,
  Inc., 2019.

\bibitem{NEURIPS2020_c5d73680}
J.~Kong, J.~Kim, and J.~Bae, ``Hifi-gan: Generative adversarial networks for
  efficient and high fidelity speech synthesis,'' in \emph{Advances in Neural
  Information Processing Systems}, H.~Larochelle, M.~Ranzato, R.~Hadsell, M.~F.
  Balcan, and H.~Lin, Eds., vol.~33.\hskip 1em plus 0.5em minus 0.4em\relax
  Curran Associates, Inc., 2020, pp. 17\,022--17\,033.

\bibitem{4815260}
F.~{Wang}, C.~{Chi}, T.~{Chan}, and Y.~{Wang}, ``Nonnegative least-correlated
  component analysis for separation of dependent sources by volume
  maximization,'' \emph{IEEE Transactions on Pattern Analysis and Machine
  Intelligence}, vol.~32, no.~5, pp. 875--888, 2010.

\bibitem{4685898}
C.~H.~Q. {Ding}, T.~{Li}, and M.~I. {Jordan}, ``Convex and semi-nonnegative
  matrix factorizations,'' \emph{IEEE Transactions on Pattern Analysis and
  Machine Intelligence}, vol.~32, no.~1, pp. 45--55, 2010.

\bibitem{7310882}
E.~{Hosseini-Asl}, J.~M. {Zurada}, and O.~{Nasraoui}, ``Deep learning of
  part-based representation of data using sparse autoencoders with
  nonnegativity constraints,'' \emph{IEEE Transactions on Neural Networks and
  Learning Systems}, vol.~27, no.~12, pp. 2486--2498, 2016.

\bibitem{7952123}
P.~{Smaragdis} and S.~{Venkataramani}, ``A neural network alternative to
  non-negative audio models,'' in \emph{2017 IEEE International Conference on
  Acoustics, Speech and Signal Processing (ICASSP)}, 2017, pp. 86--90.

\bibitem{8462116}
Y.~{Luo} and N.~{Mesgarani}, ``Tasnet: Time-domain audio separation network for
  real-time, single-channel speech separation,'' in \emph{2018 IEEE
  International Conference on Acoustics, Speech and Signal Processing
  (ICASSP)}, 2018, pp. 696--700.

\bibitem{8707065}
Y.~Luo and N.~{Mesgarani}, ``Conv-tasnet: Surpassing ideal time–frequency
  magnitude masking for speech separation,'' \emph{IEEE/ACM Transactions on
  Audio, Speech, and Language Processing}, vol.~27, no.~8, pp. 1256--1266,
  2019.

\bibitem{10.5555/3104322.3104425}
V.~Nair and G.~E. Hinton, ``Rectified linear units improve restricted boltzmann
  machines,'' in \emph{Proceedings of the 27th International Conference on
  International Conference on Machine Learning}, ser. ICML'10.\hskip 1em plus
  0.5em minus 0.4em\relax Madison, WI, USA: Omnipress, 2010, p. 807–814.

\bibitem{pmlr-v37-ioffe15}
S.~Ioffe and C.~Szegedy, ``Batch normalization: Accelerating deep network
  training by reducing internal covariate shift,'' in \emph{Proceedings of the
  32nd International Conference on Machine Learning}, ser. Proceedings of
  Machine Learning Research, F.~Bach and D.~Blei, Eds., vol.~37.\hskip 1em plus
  0.5em minus 0.4em\relax Lille, France: PMLR, 07--09 Jul 2015, pp. 448--456.

\bibitem{9053795}
R.~{Yamamoto}, E.~{Song}, and J.~M. {Kim}, ``Parallel wavegan: A fast waveform
  generation model based on generative adversarial networks with
  multi-resolution spectrogram,'' in \emph{ICASSP 2020 - 2020 IEEE
  International Conference on Acoustics, Speech and Signal Processing
  (ICASSP)}, 2020, pp. 6199--6203.

\bibitem{NIPS2014_5ca3e9b1}
I.~Goodfellow, J.~Pouget-Abadie, M.~Mirza, B.~Xu, D.~Warde-Farley, S.~Ozair,
  A.~Courville, and Y.~Bengio, ``Generative adversarial nets,'' in
  \emph{Advances in Neural Information Processing Systems}, Z.~Ghahramani,
  M.~Welling, C.~Cortes, N.~Lawrence, and K.~Q. Weinberger, Eds.,
  vol.~27.\hskip 1em plus 0.5em minus 0.4em\relax Curran Associates, Inc.,
  2014.

\bibitem{jang2021universal}
W.~Jang, D.~Lim, and J.~Yoon, ``Universal melgan: A robust neural vocoder for
  high-fidelity waveform generation in multiple domains,'' 2021.

\bibitem{yang2020multiband}
G.~Yang, S.~Yang, K.~Liu, P.~Fang, W.~Chen, and L.~Xie, ``Multi-band melgan:
  Faster waveform generation for high-quality text-to-speech,'' 2020.

\bibitem{DBLP:journals/corr/KingmaB14}
D.~P. Kingma and J.~Ba, ``Adam: {A} method for stochastic optimization,'' in
  \emph{3rd International Conference on Learning Representations, {ICLR} 2015,
  San Diego, CA, USA, May 7-9, 2015, Conference Track Proceedings}, Y.~Bengio
  and Y.~LeCun, Eds., 2015.

\bibitem{8461368}
J.~{Shen}, R.~{Pang}, R.~J. {Weiss}, M.~{Schuster}, N.~{Jaitly}, Z.~{Yang},
  Z.~{Chen}, Y.~{Zhang}, Y.~{Wang}, R.~{Skerrv-Ryan}, R.~A. {Saurous},
  Y.~{Agiomvrgiannakis}, and Y.~{Wu}, ``Natural tts synthesis by conditioning
  wavenet on mel spectrogram predictions,'' in \emph{2018 IEEE International
  Conference on Acoustics, Speech and Signal Processing (ICASSP)}, 2018, pp.
  4779--4783.

\end{thebibliography}

\end{document}